\def\PRL#1{{\sl Phys. Rev. Lett.} {\bf #1}}
\def\PRD#1{{\sl Phys. Rev.} {\bf D#1}}
\def\NPB#1{{\sl Nuc. Phys.} {\bf B#1}}
\def\PLB#1{{\sl Phys. Lett.} {\bf B#1}}

\documentstyle[12pt]{article}
\title{Bubble wall dynamics, generalised Yukawa couplings and adequate
electroweak baryogenesis in two-Higgs-doublet model}
\author{{Susmita Bhowmik Duari}\\
and \\
U. A. Yajnik\\
Department of Physics, Indian Institute of Technology,\\ Bombay
400\thinspace 076}
\date{}
\begin{document}
\maketitle
\begin{abstract}
Baryogenesis at the electroweak scale depends on  divers but
identifiable details of bubble wall dynamics and the particle
physics. We show that inclusion of the dynamics of relative phase
in two-Higgs-doublet model (2HDM) enhances the adiabatic order of
the mechanism proposed by McLerran-Shaposhnikov-Turok-Voloshin where
the scalar-scalar-vector triangle diagram with
top quark in the loop gives rise to a significant
contribution to the effective chemical potential biasing the
Chern-Simons number.
We also show that in 2HDM with less stringent constraints on
Yukawa couplings than those imposed by natural flavour conservation,
there are additional diagrams contributing
to the effective chemical potential. These two effects can combine
with several others to produce adequate baryon asymmetry at
the electroweak scale.
\end{abstract}

\section{Introduction}
The main motivation for the models of CP violation has been to
explain the $\epsilon'/\epsilon $ parameter for neutral kaon decay
simultaneously preserving
the vanishing smallness of the electric dipole moment of the neutron.
The idea that CP violation can be accomodated in extensions of the
Higgs sector
was put forward by Weinberg \cite{Wein1} in his three Higgs doublet
model, and T.D. Lee \cite{Lee1} in his two Higgs doublet model (2HDM)
with spontaneous CP violation. In these models the Yukawa couplings
have to be constrained to suppress flavor changing neutral currents.
A natural condition is to require that either $\phi_1$ couples to the
up quarks and $\phi_2$
to the down quarks or $\phi_1$ couples to all the quarks and $\phi_2$
does not couple at all. A discrete symmetry $\phi_1 \longrightarrow
-\phi_1$ is imposed on the Higgs potential to ensure suppression. In
majority opinion the minimal standard model suffers from inadequate
CP violation coming from the CKM matrix to be considered a viable
model for baryogenesis at the electroweak scale. So the 2HDM with
with CP violation in the Higgs sector has attracted attention for the
past several years in the context of electroweak baryogenesis.

McLerran-Shaposhnikov-Turok and Voloshin \cite{MacL} had shown that
in this model, the triangle diagram with one gauge boson leg and two
scalar legs and top quark in the loop contributes a correction to the
effective action $\sim {\cal O} N_{cs}$, where $\cal O$ is an operator
that can be found out by explicit evaluation of the diagram and
$N_{CS}$ is the Chern-Simons (CS) number. This correction biases the
CS number and since $\bigtriangleup N_B = n_{f} \bigtriangleup
N_{CS}$, nett baryon number results, with the sign determined by the
sign of $<\cal O >$.

Recently Yue-Liang-Wu \cite{Wu} has generalised the model of Glashow
and Weinberg \cite{Wein2} and has proposed that the value of
$\epsilon'/\epsilon $ and the limits on EDM can be well explained
with less stringent conditions on the Yukawa couplings. The Higgs
boson-fermion couplings are now much more general and no discrete
symmetries are imposed on the Higgs potential. In this class of
models fermions acquire mass from both $\phi_1$ and $\phi_2$.
Consequently in addition to the triangle diagrams with $\phi_1$
on the external legs, diagrams with  $\phi_2$ also on the external legs
start contributing to the effective action.

The MSTV proposal has been criticised on the grounds that it is a higher
order adiabatic effect, of the order $(\langle \phi \rangle^T/T)^4$.
In their considerations the relative phase between the two Higgs vacuum
expectation values was treated constant throughout bubble evolution.
Here we provide the details of the argument \cite{sbdanduay2} that the
inclusion of the dynamics of the
relative phase removes the adiabatic suppression,
making it an effect
of the order $(\langle \phi \rangle^T/T)^2$. We have also previously shown
\cite{sbdanduay1} that the uncertainties
of physical parameters of bubble formation \cite{Dineetal} can be
circumvented if we consider a string induced phase transition
\cite{Yajnik}. This is the scenario we shall be considering here as
well. It may be noted that since according to \cite{sbdanduay1}, the
string induced bubbles provide adiabatic conditions, all B-genesis
mechanisms relying on such conditions are workable. In particular the
mechanisms of Cohen and Kaplan \cite{Cohen2} and Cohen-Kaplan-Nelson
\cite{Cohen3} can also proceed through string induced bubbles.
We recapitulate  here the corresponding bubble
profile ansatz including the time evolution of the relative phase and
calculate the resultant contribution to the effective action.
Putting together the two sources of enhancement, viz., dynamics of the
relative phase and additional triangle diagrams,
we show that the mechanism has sufficient potential to give adequate
baryogenesis.

\section{The triangle diagram and correction to the effective action}
It was first pointed out by Turok and Zadrony \cite{Turok} and then
explicitly calculated by McLerran-Shaposhnikov-Turok and Voloshin
\cite{MacL}, that for 2HDM a term biasing the Chern-Simons number with
a CP odd chemical potential is contributed by
the triangle diagram of figure 1. This contribution to the effctive
action at finite temperature is
\begin{equation}
\Delta S={-7\over{4}}{\zeta(3)}{\left(m_{t}\over{\pi T}\right)^2}
     { g\over 16{{\pi }^2}}
 {1\over{{v_1}^2}}
 \times \int  (
     {\cal D}_{i}\phi_{1}^{\dagger}\sigma^{a}{\cal D}_{0}\phi_1 +
     {\cal D}_{0}\phi_{1}^\dagger \sigma^{a}{\cal D}_{i}\phi_1)
     \epsilon^{ijk} F_{jk}^{a}d^{4}x
\end{equation}
     where $m_{t}$ is the mass of the top quark, $\zeta $ is the Riemann Zeta
     function, and $\sigma^{a}$ are the Pauli matrices.
     For homogeneous but time varying configurations of the Higgs fields
     and in the ${A_0}^{a}=0$ gauge,
     \begin{eqnarray}
\Delta S_1 &=& {-i7\over{4}}{\zeta(3)}{\left(m_{t}\over{\pi T}\right)^2}
     {2\over{{v_1}^2}}
\int dt[{\phi_1}^\dagger{\cal D}_{0}\phi_{1} -
     {\cal D}_{0}\phi_1^\dagger \phi_{1}] N_{CS}\\
    & =& {\cal O}_{1} N_{CS}
     \end{eqnarray}
Clearly, here the top quark acquires mass only from $\phi_1$ as
demanded by the Glashow-Weinberg natural flavor conservation (NFC)
criteria. In Wu's model, the FC criterion is satisfied
under relaxed conditions on the Higgs-fermions couplings. In this
case the general Yukawa interactions can be written as
\begin{equation}
L_Y = \bar{q_{iL}} {(\Gamma_{D}^{a})}_{ij} D_{jR} \phi_a
+\bar{q_{iL}} {(\Gamma_{U}^{a})}_{ij} U_{jR} \bar{\phi_a}
+ \bar{l_{iL}} {(\Gamma_{E}^{a})}_{ij} E_{jR} \phi_a + h.c
\end{equation}
where $q_{i}$, $l_i$ and$\phi_a$ are $SU(2)_L$ doublet quarks,
leptons and Higgs bosons, $U_i$, $D_i$, $E_i$ are $SU(2)_L$ singlets.
$i = 1,2,...n $ is a generation label and $a = 1,2$ is a Higgs
doublet label. ${\Gamma^a}_F$ $(F = U, D, E)$ are the Yukawa coupling
matrices.
According to Wu the Glashow-Weinberg criteria can be replaced by a
theorem which states that the flavor conservation for the neutral
currents is natural in the Higgs sector or equivalently, the matrices
${\Gamma^a}_F$ $(F = U, D, E)$ are diagonalizable simultaneously by a
biunitary or biorthogonal transformation, if and only if the square
$n \times n$ ${\Gamma^a}_F$ are represented in terms of the linear
combinations of a complete set of $n \times n$ matrices $
({\Omega^\alpha}_F,  \alpha = 1,2,...n)$.
\begin{equation}
 {\Gamma^a}_F = \sum_{\alpha} {g^F}_{a\alpha} {\Omega_F}^\alpha
\end{equation}
where, ${\Omega_F}^\alpha$ satisfy the following orthogonal condition
${\Omega_F}^\alpha {({\Omega_F}^\beta)}^\dagger = {L^\alpha}_F
\delta_{\alpha\beta}$,\\ ${({\Omega_F}^\alpha)}^\dagger
{\Omega_F}^\beta= {R^\alpha}_F \delta_{\alpha\beta}$ , with the
normalization $\sum_\alpha {L^\alpha}_F = \sum_\alpha {R^\alpha}_F =
1$.
This generalised Yukawa coupling prompts us to consider another set
of triangle diagrams as shown in fig-2. The contribution at
finite temperature from these diagrams to the effective action can
similarly be calculated to be
\begin{equation}
\Delta S_2 = {-i7\over{4}}{\zeta(3)}{\left(1\over{\pi T}\right)^2}
     {{\Gamma}^{2}}^2 \int dt[{\phi_2}^\dagger({\cal D}_{0}\phi_{2}) -
     ({\cal D}_{0}\phi_2)^\dagger \phi_{2}] N_{CS}
\end{equation}
\begin{equation}
 \Delta S_3 = {-i7\over{4}}{\zeta(3)}{\left(1\over{\pi T}\right)^2}
     {\Gamma}^{1}{\Gamma}^{2}\int dt[{\phi_1}^\dagger({\cal
     D}_{0}\phi_{2}) -
     ({\cal D}_{0}\phi_1)^\dagger \phi_{2}] N_{CS}
\end{equation}
where $\Gamma_{1(2)}$ is the Yukawa coupling when the fermions couple
to $\phi_{1(2)}$.

\subsection{The bubble profile}
Now to find the bubble profile we use the following ansatze for the
finite temperature vacuum expectation value of the Higgs fields
 \begin{eqnarray}
 {\phi_{1}}^0 &=& {\rho}_{1}(r,t) e^{-i\theta(t)}  \\
 {\phi_{2}}^0 &=& {\rho}_{2}(r,t) e^{i\omega(t)}
 \end{eqnarray}
where, as pointed out by Cohen-Kaplan-Nelson \cite{Cohen1} we can
fix the unitary gauge ensuring $\theta$ is the physical pseudo-scalar
orthogonal to the Goldstone boson eaten by Z.
This gauge fixing gives the relation between $\omega$ and $\theta$
to be,
\begin{equation}
\partial_\mu \omega = {(\rho_1 / \rho_2)}^2 \partial_\mu \theta
\end{equation}
     We assume that the phase transition takes place when a combination of
$\rho_{1}$ and
     $\rho_{2}$ becomes massless in a particular direction $\gamma $ in
     the ${\phi_{1}}^0 - {\phi_{2}}^0 $ plane. Hence, in the bubble
     profile we may take
     \begin{equation}
 {\rho}_1(r,t)=\rho(r,t)\cos{\gamma} \hskip 1true in
\rho_2(r,t)= {\rho}(r,t)\sin{\gamma}
\end{equation}
         with this parametrization and finite temperature corrections,
the effective potential for the $\rho$, $\theta$ and $\omega$ is
 \begin{eqnarray}
 V_{T}(\rho,\theta,\omega)&=&{M_1}^{2}(T){\rho}^2
- ET{\rho}^{3} + {K_{1} \over {4}} {\rho}^4 \nonumber \\
&&+ {M_2}^{2}(T)(\cos{\xi} \cos{(\omega + \theta)}
+ {C_1}\sin{\xi}\sin{(\theta + \omega)} ){\rho}^2 \nonumber \\
&&+ {K_{2}\over {4} }{\rho}^4 (\cos^2{(\theta + \omega)}
     +{C_{1}} \sin^{2}{(\theta + \omega )})
 \end{eqnarray}
     where, ${M_{1}}^2$ and ${M_{2}}^2$ are temperature corrected mass
parameters and $K_{1}$ and $K_{2}$ are combinations of quartic coupling
constants with small temperature dependent corrections.  The constant
$C_{1}$ is the ratio of ${\lambda _5}$ and $\lambda _{6}$ in the
standard parameterisation of the 2HDM \cite{hhguide}, and $\xi$ is
the phase of the neutral component of zero temperature vacuum
expectation value of $\phi_2$.
     Rescaling ${\rho \longrightarrow ({2ET/{K_{1}}}){\rho}^{\prime}},
     r \longrightarrow{{r^\prime}/{2ET\over {K_{1}}}}$ and $t$
     $\longrightarrow{{t^\prime} /{2ET\over {K_{1}}}}$
  omitting the primes the potential can be written as,
  \begin{eqnarray}
     V_{T}(\rho,\theta, \omega)&=&{\left( {2ET}\over {K_{1}}\right)}^{4}
     [{K_{1}\over{2}}(1+E_{1}){\rho}^{2}
     - {K_{1}\over {2}}{\rho}^{3}+ {K_{1}\over {4}}
     {\rho}^{4} \nonumber \\
   &&+ {K_{2}\over{2}}(1+E_{2})(\cos{\xi} \cos(\omega + \theta)+
     {C_1}\sin\xi \sin(\theta + \omega) ){\rho}^{2} \nonumber \\
    &&+ {K_{2}\over {4} }{\rho}^4 (\cos^2(\theta + \omega) +
     {C_{1}} \sin^{2}(\theta + \omega ))]
  \end{eqnarray}
    where $E_{1} = \frac{M_{1}^{2}K_{1}}{2 E^{2} T^{2}}-1 $
     and $E_{2} =  \frac{M_{2}^{2}K_{1}}{2 E^{2} T^{2}}-1 $.
The geometry of string induced bubbles is cylindrical. With this in mind,
the time dependence of $\rho$ and $\theta$ can
be found by solving the following equations,
   \begin{eqnarray}
\lefteqn{ {\frac{{\partial}^{2}\rho}{\partial t^{2}}} -
     {\frac{{\partial}^{2}\rho}{\partial r^{2}}}
     -\frac{1}{r}\frac{\partial\rho}{\partial r}
     + \rho{\left( {\partial \theta}\over
     {\partial t} \right)}^2 - {3\over2}K_{1}{\rho}^2
    + K_1{\rho}^3+  {K_{1}\over 2}(1+E_{1})\rho } \nonumber\\
&&  +K_{2}C_{1}
     \sin^{2}(\theta + \xi){{\rho}^3}
  +  {K_{1}\over 2}(1+E_{2}) (\cos{\theta} +  C_{1}
     \sin{\xi} \sin{(\xi + \theta}))\rho = 0
    \end{eqnarray}
    \begin{eqnarray}
\lefteqn{ \frac{\partial^2\theta}{\partial t^2} + \frac{K_2}{4}
     C\sin{2(\theta+\xi)}{\rho^2} } \nonumber\\
&&+\frac{K_1}{2}(1+E_2)(C \sin\xi\cos{(\theta + \xi)}
-\sin(\theta+\omega-\xi))  = 0
     \end{eqnarray}
The time independent solution of $\rho$ can be found by imposing the
boundary condition $\rho \longrightarrow 0$ as $r \longrightarrow
\infty $ and ${{\partial \rho}\over{\partial r}} \longrightarrow
0$ as $ r \longrightarrow 0$. Subsequently, as the nontrivial minimum
becomes favorable, the same solutions begin to evolve in time.
The parameters used in the equations are,

\noindent
\centerline{\vbox{\noindent $K_{1}$, $K_{2}$ = 0.1 -- $0.001$, \\
     $C_{1} = 1.1$, $C = C_{1}-1$, $E_{1} = -0.074$, $E_{2} = -0.07$,\\
     $\xi = 0.2$, $E\sim 0.01$}}
The time evolution of the bubble profile $\rho(r,t)$
has been reported earlier \cite{sbdanduay1}.
In fig-3, we show the time evolution of the relative phase in the
regions where $\rho$ has become nonzero. If the initial reference value
is zero, it oscillates to reach the stationary value dictated by the
2HDM effective potential at the relevant temperature. We assume this
value to be $O(1)$ since no natural reasons prevent it from being so.

\subsection{Evaluation of the operator}
Now we can use these solutions to evaluate the average value of the
operators as,
\begin{equation}
{\cal O}_1 + {\cal O}_2 =  28 {\zeta(3)}{\left(1\over{\pi }\right)^2}
     { E^{2} \over {K_{1}^2}}A_{1} {\int {{\rho}^2}{\frac{\partial
     \theta}{\partial t}} dt}
         \end{equation}
and,
\begin{equation}
{\cal O}_3 =  28 {\zeta(3)}{\left(1\over{\pi }\right)^2}
  { E^{2} \over {K_{1}^2}}A_{2} {\int {{\rho}^2}{\frac{\partial
     \theta}{\partial t}}{e^{i(\theta + \omega)} dt}}
\end{equation}
where $A_1 = {({\Gamma_1}\cos\gamma)}^2 + {({\Gamma_2}\sin\gamma)}^2$
and $A_2 =\Gamma_1\Gamma_2 sin2\gamma$. These parameters
have to be determined from the phenomenology of the 2HDM which is yet
far from being tested by current experiments.
Knowing the solutions to eqn.s (14)-(15), we can estimate the integral
in eqn. (16) to be ${\rho_{\infty}}^2\Delta\theta$ where
$\Delta\theta$ is the nett
change in the relative phase $\theta$ at any given point as the bubble
wall sweeps past it, and $\rho_{\infty}=1$.
There is an additional contribution from the transient part of
$\theta$, which can also be calculated numerically, but is not significant.
As for the term ${\cal O}_3$, its CP odd part has the magnitude
$\rho_{\infty}^2\Delta(\sin\theta)$ from arguments already given.
Putting in other known factors, we see that
\begin{equation}
{\cal O}_1 + {\cal O}_2 \simeq\  A_1\left(E/K_1 \right)^2 \Delta\theta
\end{equation}
and a similar contribution from ${\cal O}_3$.
Recall that $E$ is the dimensionless cubic
self-coupling induced by thermal loops, and $K_1$ involves the quartic
self-couplings of the 2HDM. For naturalness we would like
$\Delta\theta$ to be $O(1)$ but the remaining factor
is numerically a small magnitude,
perhaps between $1$ and $10^{-4}$. Note however that the effect is
not suppressed by
the physics of the process viz., the bubble wall dynamics. We emphasise
again that this is the consequence of the dynamics of the relative
phase of the 2HDM.

\section{Estimation of the asymmetry}
To estimate the baryon asymmetry we assume the presence of high
temperature sphaleron processes inside
the bubble wall. The rate of such transitions per unit time per
unit volume is of the order $\sim \kappa{\alpha_{w}T}^4$, $\kappa
\sim 1$ \cite{Amb}. The number of fermions
created per unit time in the bubble wall is given by
     \begin{equation}
     B = \kappa {(\alpha _{w} T)}^{4} l S\times
{\frac {1}{T}}{\frac {{\cal O}} {l}}
     \end{equation}
where we have made use of a well established master formula
\cite{BocShap} \cite{Cohen1}, and
where $l$ and $S$ are the thickness and the surface area of the bubble
wall respectively. From which we get the baryon to photon ratio to be
     \begin{eqnarray}
     \bigtriangleup  \equiv {\frac {n_{B}}{s}}& \simeq& {\frac{1}{N_{eff}}}
     {(\alpha_{w})}^4 {\cal O } \nonumber \\
     & \simeq & 10^{-8}\times \left( \frac{E}{K_1} \right)^2 \Delta\theta
\end{eqnarray}
 where we have used $\alpha_w$ $\sim$ $10^{-\frac{3}{2}}$ and $N_{eff} \sim
100$. This answer easily accomodates the observed value of this number.

It is worth emphasising the physics of this answer which is fairly
robust against changes in the specific particle physics models. The
thermal rate contributes $10^{-6}$ through $\alpha_w^4$, and another
$10^{-2}$ is contributed by $N_{eff}$. The remaining smallness of the
answer follows from smallness of the thermal vacuum value $\langle
\phi \rangle^T$ $\sim ET/K_1$, which is small on the scale set by
$T_c$.  The appearance of this particular physical quantity has to do
with our picture of the process occuring in bubble walls as the phase
transition is yet in progress.

\section{Conclusion}
The MSTV proposal has been
criticised \cite{Cohen1} on the grounds that the operator ${\cal O}$
is of the order of $(\phi/T)^4$. Since $\phi$, i.e., the temperature
dependent vacuum value $\langle \phi \rangle^T$ is $\ll T$,
the effect was
thought to be unacceptably small. But with time variation of the
relative phase allowed, we find
the relevent operator is of the order of $(\phi/T)^2$.
The effect is therefore not intrinsically suppressed.
We also note that
the same analysis could be fruitfully applied to the proposals of
\cite{Cohen2} and \cite{Cohen3}.
Secondly, we have shown that in principle there are more diagrams
in the 2HDM contributing to the effective action. Although in making
numerical estimates we are hampered by large number of unknown
parameters in the 2HDM,
it is worth  remembering the existence of these effects
as potential sources of enhancement in electroweak baryogenesis.

\section{Acknowledgment}
This work has been supported in part by the Department of Science and
Technology, Government of India.

\newpage
\section*{Figure captions}
\begin{itemize}
\item Fig-1 The contribution to $S^{eff}$ from one higgs.
\item Fig-2 Additional diagrams contributing to $S^{eff}$ for generalised
yukawa
couplings.
\item Fig-3 Evolution of the relative phase during passage of the bubble wall.
\end{itemize}
\end{document}